\documentclass[10pt,conference]{IEEEtran}
\IEEEoverridecommandlockouts
\usepackage{cite}
\usepackage{amsmath,amssymb,amsfonts}
\usepackage{algorithmic}
\usepackage{graphicx}
\usepackage{textcomp}
\usepackage{xcolor}
\usepackage{epsfig}
\usepackage{epstopdf}
\usepackage{balance}
\usepackage{subfig}
\usepackage[printonlyused]{acronym}

\usepackage[hyperindex,breaklinks,hidelinks]{hyperref}
\usepackage{xurl}

\usepackage{breakurl}

\acrodef{fh}[FH]{frequency hopping}

\def\BibTeX{{\rm B\kern-.05em{\sc i\kern-.025em b}\kern-.08em
    T\kern-.1667em\lower.7ex\hbox{E}\kern-.125emX}}

\begin{document}

\title{A Novel LFM Waveform for Terahertz-Band Joint Radar and Communications over \\ Inter-Satellite Links}

\author{\IEEEauthorblockN{Gizem Sümen\IEEEauthorrefmark{1}\IEEEauthorrefmark{2}, Güneş Karabulut Kurt\IEEEauthorrefmark{4}, Ali Görçin\IEEEauthorrefmark{1}\IEEEauthorrefmark{3}}
\IEEEauthorblockA{\IEEEauthorrefmark{1}HİSAR Lab. @Informatics and Information Security Research Center (B{\.{I}}LGEM), T{\"{U}}B{\.{I}}TAK, Kocaeli, Turkey}
\IEEEauthorblockA{\IEEEauthorrefmark{2}Department of Electronics and Communication Engineering, Istanbul Technical University, {\.{I}}stanbul, Turkey} 
\IEEEauthorblockA{\IEEEauthorrefmark{4}Poly-Grames Research Center, Department of Electrical Engineering, Polytechnique Montréal, Montréal, QC, Canada}
\IEEEauthorblockA{\IEEEauthorrefmark{3}Department of Electronics and Communication Engineering, Yıldız Technical University, {\.{I}}stanbul, Turkey} \\

Emails: \texttt{gizem.sumen@tubitak.gov.tr, gunes.kurt@polymtl.ca, agorcin@yildiz.edu.tr}
}
\maketitle
\begin{abstract}
There is no doubt that we need to keep our eyes on the sky as satellite networks aim to address the demands of 6G and beyond communications systems. On the other hand, the existence of millions of space debris pieces, large or small,  poses a threat to the new space communications systems which consist of large number of small satellites, especially in the low-orbit. In this study, a dual-functioning pulsed linear frequency modulated (LFM) waveform at Terahertz (THz) bands is proposed for both wireless communications and space debris sensing over low-orbit inter-satellite links (ISLs). Initially, the ambiguity function of the proposed waveform is derived. Then, velocity and range estimation performance for the radar function and bit error rate performance for the communications function are investigated. Simulation results indicate significant performance gains in the THz-bands compared to the legacy LFM systems.
\end{abstract}

\section{Introduction}
With the vast availability of satellite production materials and due to the drastic cost-reduction related prevalence of launching mechanisms, satellite networks have become one of the essential pieces of the puzzle in providing 6G demands such as extending wideband Internet access over urban, semi-urban, and remote rural areas with low latency and reliability. On the other hand, with the acceleration of space studies, a problem that has been on the agenda for a long time became more apparent; space debris. According to the latest data from the European Space Agency, 2900 of the 8300 satellites in space are currently not functioning~\cite{space}. Moreover, the number of 10cm to 1mm space debris objects has been identified as approximately 131 million, and more than 630 events have been reported where these objects collided with operating systems and resulted in fragmentation. In order to avoid financial and operational damages caused by space debris, ground-based radar systems has been used for a long time. However, small-scale objects cannot be detected in ground-based radars, as resolution trade-off has to be made to reach a high range. In June 2021, the 5mm hole opened by the impact of space debris on the robotic arm on duty at the International Space Station~\cite{magazine_2021}. 

Such scenarios are expected to become more prevalent in the near future, thus developing space debris radar that operates in Terahertz (THz) frequencies becomes essential to detect millimeter-scale objects since the utilization of ground-based radars in the THz bands is not a realistic option when the distance between Earth and the debris is considered. This problem can be tackled by arming some of the low-orbit mega-constellation satellites with THz-band space debris radars, since the biggest challenge of the THz band, molecular absorption is partially eliminated at higher atmospheric layers~\cite{civas2021terahertz}, \textit{i.e.}, the signal can reach much greater distances. One of the essential features sought in satellites forming mega constellations is the weight for low deployment costs associated with launching into orbit~\cite{nie2021channel}. Therefore, adding extra radar hardware to the system will not be preferred. If inter-satellite links (ISLs) and space debris radar can be implemented with a single waveform, a win-win situation occurs in terms of weight optimization and cost-efficiency. Furthermore, space-based joint radar communication will be spectral efficient due to the fact that additional spectrum will not be used either for radar or communications. When the number of satellites is considered, spectral efficiency also stands out as one of the critical parameters for inter-satellite communications. Therefore, a joint radar and communications (JRC) system would bring essential advantages to ISLs since it would be possible to detect space debris and perform inter-satellite communication with the same waveform. 
 
\subsection{Related Work}
Space debris monitoring studies date back to the early 1990s~\cite{reynolds1990review,mehrholz1995radar,stansbery1995characterization}. In particular, observation with ground-based radars, optical telescopes and laser have been studied in depth~\cite{ klinkrad2006space,shell2010optimizing,esmiller2014space}. However, optical studies do not fully meet the need due to the problems in tracking and being sensitive to weather conditions while ground-based radars have difficulty in distinguishing small objects. In~\cite{carl1993space}, a hybrid system with space-borne microwave radar and ground-based radar is proposed to eliminate the disadvantages on both applications. Space-born radar is investigated in terms of suitable parameters for the detection of space debris in~\cite{fu2008overview}. 
The space-based radar concept required for the millimeter-scale space debris object is proposed and the potential limitations of the pulse and continuous-wave (CW) radars are discussed in~\cite{cerutti2017preliminary}. Space debris radar operating at THz frequencies has been considered in~\cite{yang2016terahertz}. Also, in~\cite{yang2017three}, high-resolution THz radar has been developed for proof-of-concept. On the other hand, JRC has become a hot topic recently, as it both reduces costs and provides spectral efficiency. In~\cite{gaglione2016fractional}, Fractional Fourier Transform based multiplexing in which data is transported with different chirp subcarriers schemes is presented. Information is embedded in the direction of the linear frequency modulated (LFM) pulse, up-chirp and down-chirp in~\cite{roberton2003integrated}. A multi functional ultra wideband (UWB) communication and radar system is  proposed~\cite{saddik2007ultra}. 
However, scenarios and suggested waveforms are generally handled by focusing on automotive radars.

~\cite{anttonen2021space} introduced the joint space debris monitoring and satellite communications concept proposing millimeter-wave cyclic prefix orthogonal frequency division multiplexing (OFDM) as the JRC waveform. However, due to its low power efficiency from high peak to average power ratio (PAPR), the utilization of OFDM in space-based JRC systems seems to be a long shot.
\subsection{Contributions}
\begin{figure}[h!]
    \centering
    \includegraphics[width=.44\textwidth]{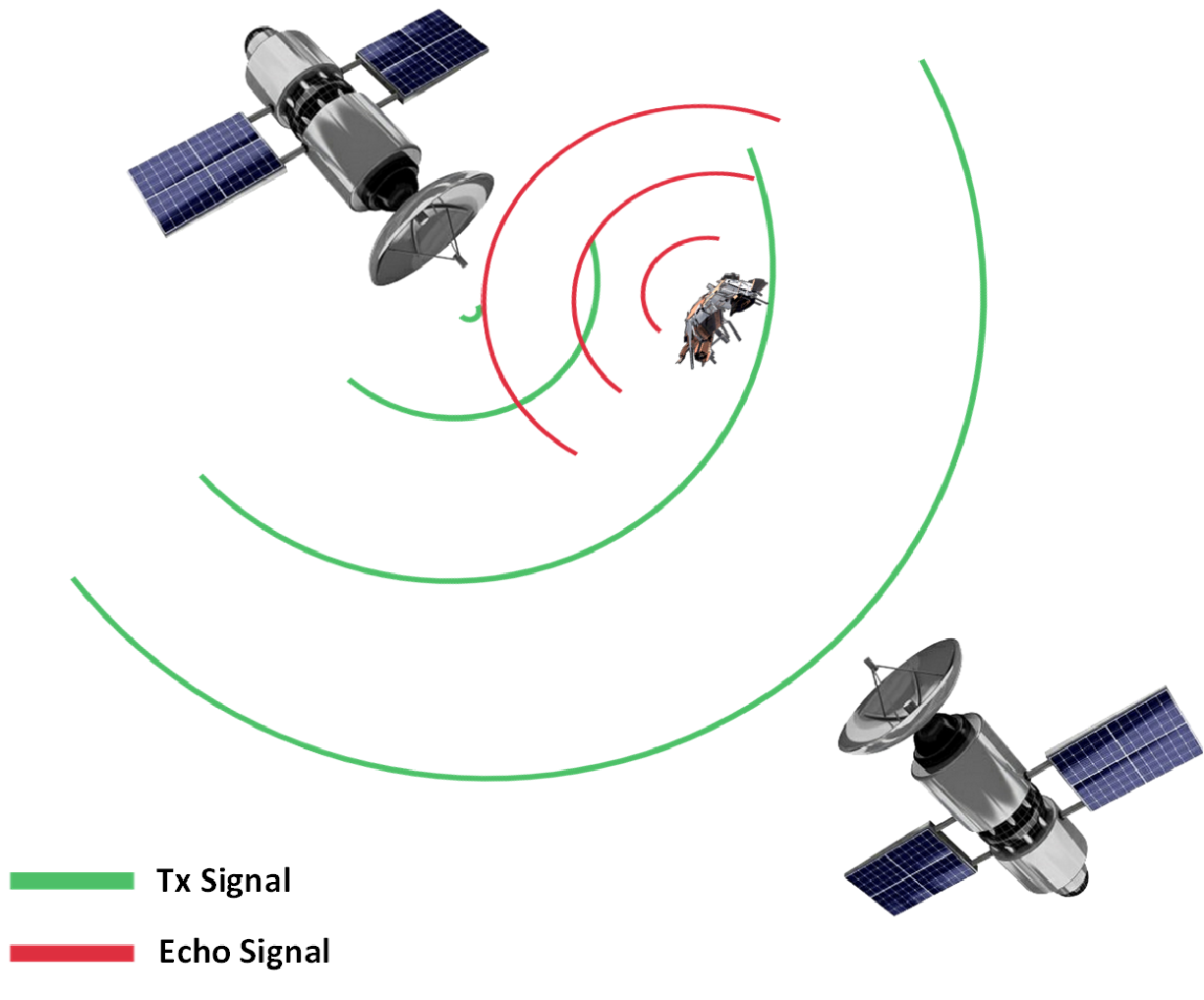}
    \caption{Proposed space-based JRC system for satellites.}
    \label{fig:1}
\end{figure}

An optimal wireless communications setting in terms of operational frequencies and waveform to achieve both space debris detection and ISLs becomes necessary as the debris problem is a growing concern. To this end, this study proposes following solutions:
\paragraph*{Contribution 1}
In our work, we consider space debris radar and ISLs as a pair as illustrated in Fig.~\ref{fig:1}. Space debris detection is performed with the echoes during inter-satellite communication. Thus, space debris can be detected in addition to realizing communication between satellites.
\paragraph*{Contribution 2}
Space-based JRC structure at THz frequencies is proposed for the first time. Since the reduced range of the THz due to water vapor molecules becomes less of a problem for low-orbit satellites, THz-band is a good candidate for ISLs~\cite{nie2021channel}. 
\paragraph*{Contribution 3}
The proposed waveform is suitable to meet the high throughput requirement of THz ISLs, as it is chosen as a chirp-based signal with relatively low PAPR. In addition, since radar and communications receivers are made of the same blocks, a two-way system can be built without incurring an extra cost. In fact, it is suitable for establishing a structure that will provide cooperative detection and communication in mega-constellations.
\paragraph*{Contribution 4}
Frequency modulated (FM) radars are known for low-cost and high-resolution measurements. As the bandwidth increases, the resolution improves. Thus, FM radar schemes are suitable for wideband THz communications. However, LFM radars suffer from the delay-Doppler coupling effect when the target is in motion. Thus, range and velocity cannot be detected unambiguously in the moving target scenario. In this study, we proposed a triangle LFM and V-shaped LFM (V-LFM) pulse as modulation schemes; while the shape of the waveform eliminates ambiguity, the frequency modulation provides a fine resolution. 

The rest of the paper is organized as follows; the system model provided in Section~\ref{sec:3}. Section~\ref{sec:2} details the proposed waveform technique and its implementation at THz frequencies. In Section~\ref{sec:4} simulation details and results are presented. Possible future work is described in Section~\ref{sec:6} along with conclusion of the study.

\section{System Description}
\label{sec:3}
\begin{figure*}[h!]
    \centering
    \includegraphics[width=0.98\textwidth]{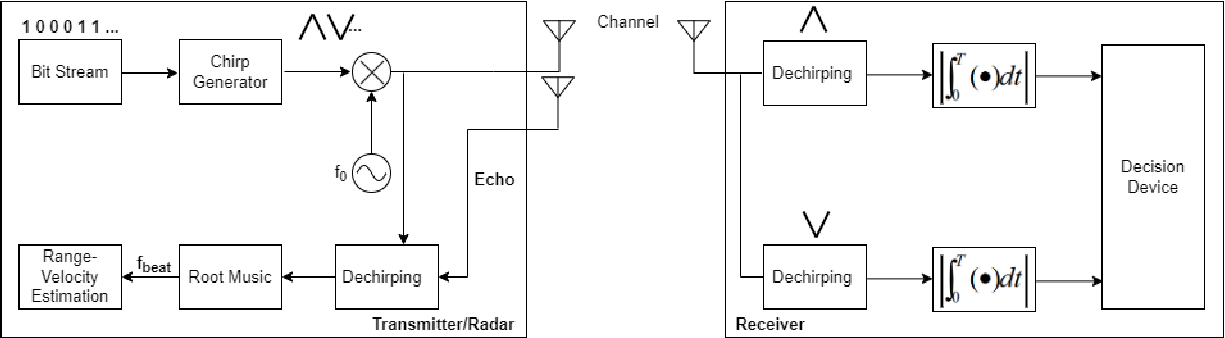}
    \caption{Block diagram of dual space debris radar and wireless communication system for communications satellites.}
    \label{fig:overall}
\end{figure*}

%This section explains how the JRC system model works for space debris detection and ISL dual scenario.  
The LFM radar signal is generated by frequency sweeping over a specific frequency range and period. FM radars are also divided into groups as frequency modulated continuous wave (FMCW) and FM pulse radar. Pulse radars send short-time pulses and calculate the distance of the target with the time delay between the sent and received signal. During the pulse radar transmitting, a strong leak occurs to the receiver~\cite{salazar2021progressive}. Since it cannot perform the receive operation during the pulse, objects at short distances cannot be detected, also called the blind spot problem. Therefore, pulse radar is not suitable for space debris detection. On the other hand, chirp signals are sent continuously in FMCW radar and echoes are received at the same time while transmitting. Since there is no blind spot problem in this case, FMCW can be used for space debris detection in the space-based JRC system. However, the diversity required for bit transfer cannot be achieved with FMCW. Therefore, in the designed JRC system, although the signal form is in the pulse form, the pulse duration is adjusted so that the echo returns before the pulse period ends. Consequently, the proposed system is FMCW radar, but from the communication perspective, it is pulse-based. The overall system model is illustrated in Fig.~\ref{fig:overall} and can be examined in three sections; transmitter/radar, receiver, and the channel.
\subsection{Transmitter/Radar}
In the transmitter part, the 0s and 1s are first converted to V-LFM pulse and triangle LFM pulse in the chirp generator block, respectively. The properties of the waveform generated in this block will be discussed in the Section~\ref{sec:2}. Inspired by~\cite{yang2017three}, the operating frequency, $f_{0}$, is selected as 340GHz since communications in this band is becoming prevalent faster than others. While the chirp duration is set to 300$\mu$s, the chirp bandwidth is selected as 288MHz.
Dechirping block in Fig.~\ref{fig:overall} is utilized in the receiver part of the radar. In FMCW radars the range and velocity are determined based on the difference in transmitted and received frequency called the beat frequency, ${f_{beat}}$. The dechirping and root-Multiple Signal Classification (root-MUSIC) blocks in the system are intended to detect this ${f_{beat}}$. In the dechirping process, the incoming echo is mixed with a copy of the transmitted chirp~\cite{liu2018dechirping}. Since the dechirping process reduces the effective intermediate frequency bandwidth, the speed requirement of the analog to digital converter (ADC) is lower than when the matched filter is used. Besides the mentioned advantages of dechirping process, the most significant disadvantage is the decrease in performance in the low SNR region compared to the matched filter~\cite{wang2011comparison}. The output of the dechirping process is given as input to, a variant of MUSIC, which is root-MUSIC algorithm in order to compensate this issue. After obtaining the beat frequency, range and velocity estimation can be done as described in Section~\ref{sec:2}.
\subsection{Receiver}
As shown in Fig.~\ref{fig:overall}, the received signal is fed into two branches and mixed with two reference waveform, separately. By using the dechirping block, a similarity is achieved between the radar receiver and the communication receiver. Thus, considering the bidirectional communication scenario, it is ensured that it does not bring an extra burden to the hardware. After the dechirping process, the samples are then integrated over a chirp period and sent to the decision device. The decision device determines which branch has the greater value, and the demodulation process is completed.
\subsection{Channel}
In this study, slightly non-line of side (NLOS) conditions are considered due to the fact that topologically some of the low-orbit satellites will not have direct LOS between them in particular cases, thus Rician fading with the shape parameter of $K = 10$ is considered for a realistic model~\cite{tekbiyik2021graph}. Furthermore, from the radar perspective, as indicated in~\cite{cerutti2017preliminary}, maximum transversal velocity of a given space debris has been estimated as 15km/s. In addition, the maximum distance of space debris to a detecting radar has been determined as 500m~\cite{cerutti2017preliminary} which is sufficient considering that there are robotic designs with a response time of less than five hundredths of a second~\cite{perrin-2014}. 
\section{the proposed method}
\label{sec:2}
\begin{figure}[tb]
    \centering
    \includegraphics[width=.49\textwidth]{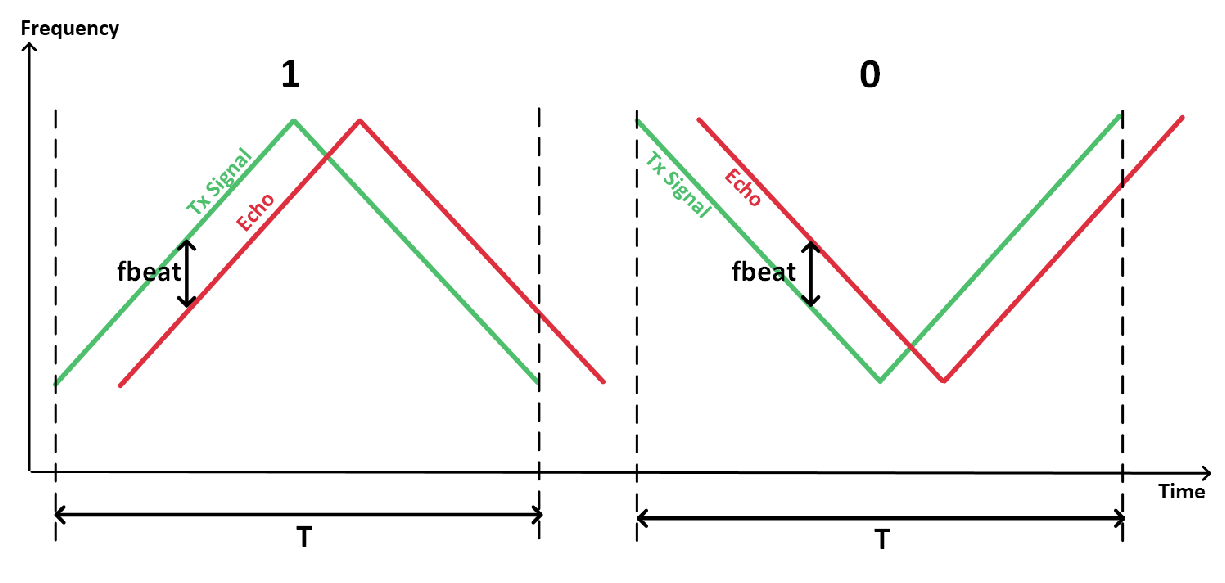}
    \caption{ Proposed dual radar communication waveform.}
    \label{fig:2}
\end{figure}

Complex baseband equivalent of received signal $r(t)$ for communications component is modeled as
\begin{equation}
r(t)=x(t) * \rho(t)+\omega(t),
\end{equation}
where $x(t)$ is the complex baseband equivalent of transmitted signal; $\rho(t)$ denotes channel impulse response; $\omega(t)$ is a sample function of additive white Gaussian noise (AWGN) process with a flat power spectral density $N_{0}/2$ W/Hz. As mentioned earlier, while the proposed waveform is in pulse form from the communication perspective, it can be analyzed as FMCW since the echo reaches the radar before the transmitted signal ends. The linear-FMCW radar signal with bandwidth $\Delta F$ and pulse duration T can be written as
\begin{equation}
s(t)=A { \exp \left(j 2 \pi\left(f_{0} t+\frac{1}{2}(-1)^{(1-c)} \mu t^{2}\right)\right)},
\end{equation}
where $\mu =\Delta F/T $, A is the envelope of chirp signal, $f_{0}$ is the center frequency and $c$ is for determination of the chirp direction. As frequency increases and decreases over time, the waveform is called \textit{up-chirp} (c=1) and \textit{down-chirp} (c=0), respectively. Echo signal received by FMCW radar can be expressed as
\begin{equation}
s_{echo}(t)=A_{r}(t) \exp \left( j 2\pi ( f_{beat} \cdot t+\phi)\right),
\end{equation}
where $A_{r}(t)$ is the amplitude term and $\phi$ is the phase term of the echo. The beat frequency ${f_{beat}}$ of up-chirp and down-chirp can be expressed as~\cite{mitsumoto1999fmcw}
\begin{equation}
f_{\text {beat,up }}=\frac{\Delta F}{T} \cdot \frac{4 R}{c}-\frac{2 V_{r}}{\lambda},
\label{eq:up}
\end{equation}
\begin{equation}
f_{\text {beat,down }}=\frac{\Delta F}{T} \cdot \frac{4 R}{c}+\frac{2 V_{r}}{\lambda},
\label{eq:down}
\end{equation}
where R is target range, $V_{r}$ is the velocity of the target, $c$ is speed of light and $\lambda$ is wavelength of the signal. As can be seen from Eq. \ref{eq:up} and  Eq. \ref{eq:down}, ${f_{beat}}$ is dependent on by both velocity and range. Therefore, due to range-Doppler coupling, ambiguity problem arises in velocity and range estimations of moving target~\cite{mitsumoto1999fmcw}. Triangle LFM and V-LFM pulse waveforms created by combining up-chirp and down-chirp signals eliminate this problem as velocity components cancel each other out in the ${f_{beat}}$ equation. The frequency-time representation of the proposed dual radar waveform is shown in Fig.~\ref{fig:2}.  0s and 1s are carried by V-LFM pulse and triangle LFM pulse, respectively.

Complex envelope of triangle LFM pulse and V-LFM pulse can be  expressed as 
\begin{equation}
\tilde{x}_{\text{tri-LFM}}(t)=\left\{\begin{array}{lr}
\frac{1}{\sqrt{2 T}} \operatorname{rect}(t / 2 T) e^{j \pi \mu t^{2}} & -T < t < 0 \\
\frac{1}{\sqrt{2 T}} \operatorname{rect}(t / 2 T) e^{-j \pi \mu t^{2}} & 0 < t < T \\
0 & \text { else }
\end{array}\right.
\label{eq:tri}
\end{equation}

 \begin{equation}
\tilde{x}_{\text{V-LFM}}(t)=\left\{\begin{array}{lr}
\frac{1}{\sqrt{2 T}} \operatorname{rect}(t / 2 T) e^{-j \pi \mu t^{2}} & -T < t < 0 \\
\frac{1}{\sqrt{2 T}} \operatorname{rect}(t / 2 T) e^{j \pi \mu t^{2}} & 0 < t < T \\
0 & \text { else }
\end{array}\right.
\label{eq:v}
\end{equation}
where $\tilde{x}_{\text{tri-LFM}}(t)$ and $\tilde{x}_{\text{V-LFM}}$ refer complex envelope of triangle LFM pulse and V-LFM pulse, respectively. Finally, when combined waveforms are used, the range and velocity can be estimated without ambiguity as in Eq.~\eqref{eq:r} and Eq.~\eqref{eq:velocity} by adding/subtracting Eq.~\eqref{eq:up} and Eq.~\eqref{eq:down}. 
\begin{equation}
    R=\frac{T c}{8 \Delta F}\left(f_{ beat,down }-f_{{beat,up }} \right)
    \label{eq:r}
\end{equation}

\begin{equation}
    V_{r}=\frac{\lambda}{4}\left(f_{ beat,down }-f_{{beat,up }} \right)
    \label{eq:velocity}
\end{equation} 

In the following section, the ambiguity function of the proposed waveform will be derived. The ambiguity function not only provides information with the range and Doppler resolution of the waveform but also allows determining which applications the waveform is suitable for. The ambiguity function of a radar signal~\cite{mahafza2005radar} can be expressed as
\begin{equation}
\left|\chi\left(\tau, f_{d}\right)\right|^{2}=\left|\int_{-\infty}^{\infty} \tilde{x}(t) \tilde{x}^{*}(t-\tau) e^{j 2 \pi f_{d} t} d t\right|^{2},
\end{equation}
where $f_{d}$ is Doppler frequency of the moving target, and $\tau$ time delay that depends on range. In order to facilitate the calculation of the ambiguity function of the proposed waveform, considering the complex envelope of up-chirp as $\tilde{x}_{1}$ and down chirp as $\tilde{x}_{2}$ complex envelope of triangle LFM can be expressed as
\begin{equation}
\tilde{x}_{tri}(t)=\tilde{x}_{1}(t)+\tilde{x}_{2}(t) \quad ;-T<t<T.
\end{equation} 
Ambiguity function of $\tilde{x}_{tri}(t)$ can be expressed as~\cite{tan2012resolution}

\begin{eqnarray*}
\chi_{tri}(\tau,f_{d})&=&\chi_{u_{11}}(\tau,f_{d})+\chi_{u_{22}}(\tau,f_{d})+\chi_{u_{12}}(\tau,f_{d})\\
&&+e^{-j2 \pi f_{d}\tau}\chi_{u_{12}}^{*}(-\tau,-f_{d}),
\end{eqnarray*}
where $\chi_{u_{11}}$ and $\chi_{u_{22}}$is self broadband ambiguity function of up-chirp and down-chirp, respectively. $\chi_{u_{12}}$ is cross broadband ambiguity function. By adapting the formulas in~\cite{tan2012resolution} to electromagnetic waves  
\begin{eqnarray*}
\chi_{u_{11} }(\tau,f_{d})&\approx&\frac{1}{4 T \sqrt{|\mu|}}{ e}^{j \pi\left(\frac{f_{d}^{2}}{2\mu }-\mu \tau^{2}\right)}\{\left[C\left(x_{2}\right)-C\left(x_{1}\right)\right]\\
& & +\operatorname{sgn}(f_{d}) \cdot j\left[S\left(x_{2}\right)-S\left(x_{1}\right)\right]\},
\end{eqnarray*}
where $\operatorname{sgn}(\cdot)$ is the sign function, $C(\cdot)$ and $S(\cdot)$  denote the cosine and sine integral. $x_{1}=2{ \sqrt{|\mu|}}[a(\tau)-f_{d} /(2 \mu )]$ and $x_{2}=2{ \sqrt{|\mu|}}[b(\tau)-f_{d} /(2 \mu )]$ where $a(\tau)$ and $b(\tau)$ can be expressed as
\begin{equation}
\left\{\begin{array}{lr}
a(\tau)=-T-\tau,b(\tau)=0 & -T<\tau<0 \\
a(\tau)=-T,b(\tau)=-\tau & 0<\tau<T
\end{array}\right.
\end{equation}

\begin{eqnarray*}
\chi_{u_{22} }(\tau,f_{d})&\approx&\frac{1}{4 T \sqrt{|\mu|}}{ e}^{-j \pi\left(\frac{f_{d}^{2}}{2\mu }-\mu \tau^{2}\right)}\{\left[C\left(x_{4}\right)-C\left(x_{3}\right)\right]\\
& & -\operatorname{sgn}(f_{d}) \cdot j\left[S\left(x_{4}\right)-S\left(x_{3}\right)\right]\},
\end{eqnarray*}
where $x_{3}=2{ \sqrt{|\mu|}}[a(\tau)+f_{d} /(2 \mu )]$ and $x_{4}=2{ \sqrt{|\mu|}}[b(\tau)+f_{d} /(2 \mu )]$. $a(\tau)$ and $b(\tau)$ can be expressed as
\begin{equation}
\left\{\begin{array}{lr}
a(\tau)=-\tau, b(\tau)=T & -T<\tau<0 \\
a(\tau)=0, b(\tau)=T-\tau & 0<\tau<T
\end{array}\right.
\end{equation}
Finally, cross broadband ambiguity function can be calculated as
\begin{eqnarray*}
\chi_{u_{12} }(\tau,f_{d})=\frac{1}{4 T \sqrt{|\mu|}} { e}^{j \pi \mu \tau^{2}-\frac{j \pi f_{d}^{2}}{2 \mu}}\{\left[C\left(x_{6}\right)-C\left(x_{5}\right)\right]\\
+j\left[S\left(x_{6}\right)-S\left(x_{5}\right)\right]\},
\end{eqnarray*}
where $x_{5}=2{ \sqrt{|\mu|}}[a(\tau)+f_{d} /(2 \mu )]$ and $x_{6}=2{ \sqrt{|\mu|}}[b(\tau)+f_{d} /(2 \mu )]$. $a(\tau)$ and $b(\tau)$ can be expressed as
\begin{equation}
\left\{\begin{array}{lr}
a(\tau)=-\tau, b(\tau)=0 & 0<\tau<T \\
a(\tau)=-T, b(\tau)=T-\tau & T<\tau<2T
\end{array}\right.
\end{equation}

%%%  VLFM için***
These results imply that the ambiguity function is dependent on $f_{0}$, $\Delta F$, and T. Range and velocity resolution can also be determined in the ambiguity function by examining the $f_{d}=0$ and $\tau=0$ states, respectively~\cite{tan2012resolution}.

\section{simulation results}
\label{sec:4}

\begin{figure}[t]
\centering
\subfloat[Range accuracy comparison between FMCW and the proposed waveform.]{%
  \includegraphics[width=.45\textwidth]{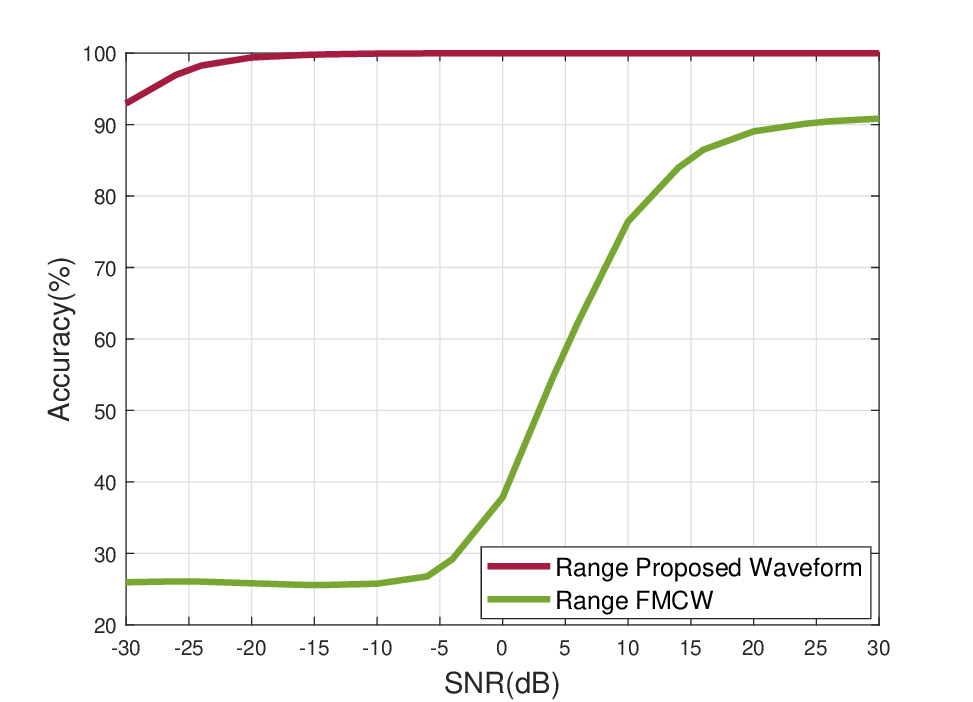}%
  \label{fig:radar_performans_range}%
}\qquad
\subfloat[Velocity accuracy comparison between FMCW and the proposed waveform.]{%
  \includegraphics[width=.45\textwidth]{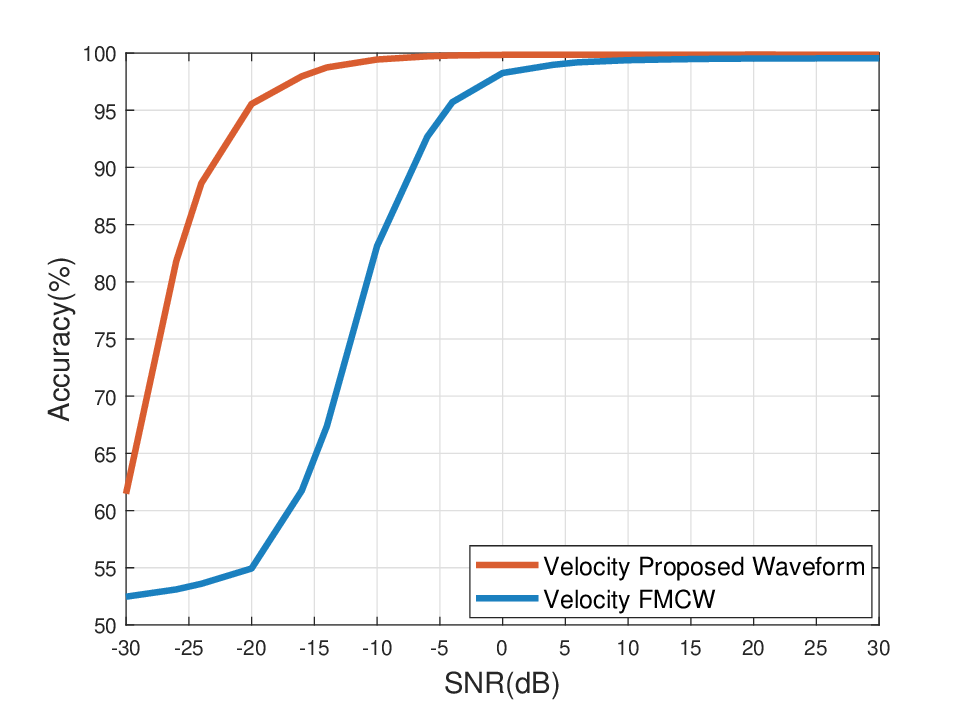}%
   \label{fig:radar_performans_vel}%
}
\caption{Simulation results for radar parameter performance of the proposed method.}
\label{fig:radar_performans}
\end{figure}

Contemporary satellite communications are carried out in L-band, Ka-band, V-band~\cite{nie2021channel}. However, due to the next generation wireless network demands, massive number of low-orbit satellites operating at millimeter waves, free space optics (FSO), and THz frequencies are being proposed for ISLs. There are quite a few studies on ISLs, but even less work from the JRC perspective which can address multiple issues jointly. Since this is the initial study examining JRC in space at THz frequencies with a LFM waveform, comparison possibilities are quite limited. Thus, in order to understand the value of the proposed waveform and its performance, especially from the radar point of view, velocity and range estimation comparisons are made with respect to the FMCW radar. The exact same radar receiver diagram in Fig.~\ref{fig:overall} is utilized for the FMCW waveform to ensure the validity of the comparison of the waveforms. The Monte Carlo simulation is carried out in MATLAB. Space debris with a constant diameter of 1mm is modeled as a with range $X_{d} \sim \mathcal{N}(250,\,70^{2})$ m and with a relative velocity $X_{v} \sim \mathcal{N}(10,\,2^{2})$ km/s. Radar performance comparison of FMCW and proposed waveform is given in the Fig.~\ref{fig:radar_performans}.

The range and velocity accuracy are defined as
\begin{equation}
    \% R_{accuracy}=100-\frac{|R-R_{estimated}|}{R} x 100 
\end{equation} 

\begin{equation}
    \% V_{accuracy}=100-\frac{|V-V_{estimated}|}{V} x 100 
\end{equation} 

The simulation results in Fig.~\ref{fig:radar_performans_range} imply that the combined use of up and down chirp eliminates the delay-Doppler coupling. The accuracy plateau value of FMCW in Fig.~\ref{fig:radar_performans_range} can be increased by decreasing the pulse duration because longer sweep time makes the range Doppler coupling more prominent for FMCW radar. However, due to ambiguity, it will never go higher than the accuracy of the proposed waveform. In velocity estimation, even though the difference in performance decreases to $0.5\%$ after 10dB, the superiority of the proposed model over the low SNR region can be seen in~\ref{fig:radar_performans_vel}.

Communication performance comparison is carried out with the bit error rate (BER) versus SNR simulation of the proposed waveform and LFM pulses (up chirp-down chirp). Carrying the bit with a chirp waveform in both waveforms actually creates a sort of channel coding effect. Therefore, SNR has been normalized according to chirp length so that the comparison between the two waveforms can be made incisively. Please note that since the satellites are assumed to be at the low-orbit region, some level of atmospheric effects will still occur thus the channel effect is modeled with Rician fading with the shape parameter $K=10$, as slightly NLOS status should be considered due to the topological reasons and maneuvers of the satellites. Also note that after dechirping the incoming signal to demodulate LFM pulse waveforms, the samples are integrated over one chirp period.
\begin{figure}[htp!]
    \centering
    \includegraphics[width=.45\textwidth]{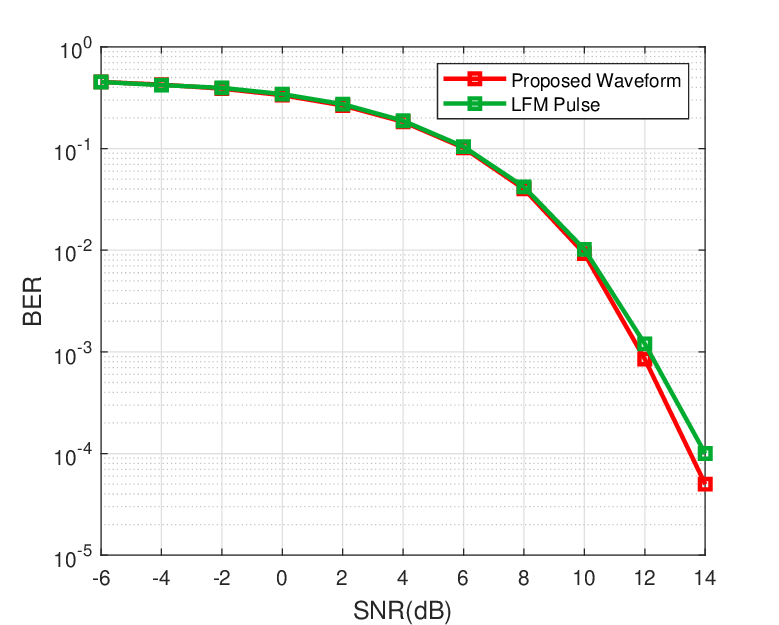}
    \caption{BER comparision of proposed waveform and LFM pulse.}
    \label{fig:ber}
\end{figure}

Fig.~\ref{fig:ber} indicates that the proposed scheme starts to improve in terms of BER performance aroud $0$ dB region and as the SNR becomes positive BER improves exponentially at every step. It can be seen that, especially after 8 dB, the proposed waveform performs better than the LFM pulse. This is due to the higher sample rate necessity of the matched filter technique~\cite{noor2022hybrid} used to demodulate the LFM pulse. Although the need for the high sample rate is compensated by the noise robustness of the matched filter at the low SNR region, the dechirping process is performed better at the high SNR region.
\section{Future Work and Conclusion}
\label{sec:6}
Space debris detection gained much more attention with the acceleration of space operations. In this study, space debris radar and THz satellite to satellite wireless communications are jointly investigated from the JRC perspective and a novel solution is proposed; radar and communications are considered to operate utilizing the same waveform without any extra requirement. In addition, while designing the waveform, space constraints and requirements such as high throughput, high resolution, low cost and low weight are taken into consideration. The superiority of the designed waveform is demonstrated via the simulation results. As indicated by the simulation results, while the effect of doppler-range coupling in FMCW is eliminated in a combination of chirp waveforms, communication is enabled with the waveform being in the form of a pulse.

Benefiting from ISLs for space debris detection is a new concept, thus, there is a lot to work to be done. For instance, cooperative communications as a distributed radar system can be considered to improve the detection of echoes coming back from space debris. In addition, spectral efficiency can be increased by designing different waveforms. Dechirping is employed herein to reduce the sample rate requirement. If sufficient sample rate can be provided, matched filter can also be considered to improve performance at the low SNR region.

\section*{Acknowledgment}
This publication is made possible in part by NPRP award [NPRP12S-0225-190152] from Qatar National Research Fund, a member of Qatar Foundation. The statements made herein are solely the responsibility of the authors.

This work was supported in part by NSERC Discovery Grant.

% % % % % % % % % % % % % % % % % % % % % % % % % % % %
\balance
\bibliographystyle{IEEEtran}
\bibliography{references.bib}
% % % % % % % % % % % % % % % % % % % % % % % % % % % %

\end{document}